\documentclass[aps,pra,reprint]{revtex4-2}
\usepackage[utf8]{inputenc}
\usepackage{tabularx}
\usepackage{graphicx}
\usepackage{dcolumn}
\usepackage{bm}
\usepackage{amsmath}
\usepackage{amssymb}
\usepackage{subeqnarray}
\usepackage{cases}
\usepackage{enumitem}
\usepackage{xcolor}
\usepackage{bbm}

\usepackage{esint}
\usepackage{subcaption}

\newcommand{\rmc}{c}%{{\rm c}}
\newcommand{\rmi}{{\rm i}}

\newcommand{\rmd}{{\rm d}}
\newcommand{\rmm}{m}%{{\rm m}}
\newcommand{\rme}{{\rm e}}
\newcommand{\rmM}{M}%{{\rm M}}

\mathchardef\NABLA"272
\newcommand*{\Nabla}{\boldsymbol\NABLA}
\let\nabla\Nabla

\usepackage[capitalise]{cleveref}
\newcommand{\barecref}[1]{%
	\unskip % remove any space token immediately before the macro
  \begingroup
    \creflabelformat{equation}{(##2##1##3)}%
    \crefname{equation}{}{}%
    \Crefname{equation}{}{}%
    \cref{#1}%
  \endgroup
}
\begin{document}

\title{
Calculating propagators for relativistic wave equations
}

\author{Mingjie Li}
\email{ml813@exeter.ac.uk}

\author{S. A. R. Horsley}
\email{S.Horsley@exeter.ac.uk}
\affiliation{Department of Physics and Astronomy, University of Exeter, Stocker Road, Exeter, UK, EX4 4QL}

\vspace{10pt}

\begin{abstract}
Here we examine the propagation of relativistic fields in spacetime using the viewpoint applied to derive the Rayleigh--Sommerfeld diffraction integral in three--dimensional space.  We use this theory to find the propagators for both the Klein--Gordon equation and the Dirac equation. Based on these results, we highlight some apparently overlooked aspects of the Feynman checkerboard model as it is usually applied to the Dirac equation.

\end{abstract}

\maketitle

\vspace{2pc}

\section{Introduction}
Inspired by Dirac's discovery of a connection between the Lagrangian in classical mechanics and Heisenberg's matrix mechanics~\cite{Dirac1933}, Feynman proposed the path integral formulation~\cite{Feynman1948} to describe the behavior of matter waves in terms of a sum over path--dependent phases, instead of a wave equation, providing an intuitive physical picture on how matter waves propagate through various paths.  This approach has proved to be useful for calculating time evolution in quantum mechanics, both in single particle theory, and field theory~\cite{junker2024path,benzair2024path,laenen2009path,hertz2016path}.  The path integral shows that there is a contribution to the phase of a matter wave that is proportional to the Hamiltonian action of the particle, with the classical path dominating in the classical limit, $\hbar\to0$.

There is a similar situation in optics, in diffraction theory, where Kirchoff's theory gives the optical wave in terms of a sum of source waves distributed across a wavefront, which can be understood as a sum over paths.  Although diffraction theory is usually applied in three--dimensional space---time being neglected through the assumption of time--harmonic fields---the same approach can be extended to spacetime.  This was first examined some time ago in e.g. Ref.~\cite{brukner1997diffraction}, where Kirchhoff diffraction theory was extended to treat propagation in $(3+1)$ dimensions.  In this work, based on this connection between the path integral and diffraction theory, we use Rayleigh diffraction theory to construct the propagators for the Klein--Gordon~\cite{KleinEqn} and Dirac equations~\cite{DiracEqn}.  We note that our work is connected to the application of diffraction theory including time--evolution, which has found wide-ranging applications in recent research~\cite{goussev2012huygens,goussev2013diffraction,barbier2022phase,jones2015path,beau2015three}.   

In what follows we shall compare the results of our approach to a different method for calculating the propagator, given by Feynman~\cite{FeynmanQM}, for the particular case of the $(1+1)$D Dirac equation.  This is known as the Feynman checkerboard model and differs from the sum over paths applied to the Schr\"{o}dinger equation~\cite{SchrodingerEqn}.  Note that this checkerboard approach has been extended to the full Dirac equation in $(3+1)$ dimensions, see e.g. Ref.~\cite{ord2023feynman}.  We shall show that, using our approach, we reclaim the checkerboard propagator but with additional terms that are non--zero only on the light cone, which are essential to reclaim the correct zero mass limit.

\section{The propagator for the Klein--Gordon equation}

\begin{figure}[htb!]
    \centering
    \includegraphics[width=0.8\linewidth]{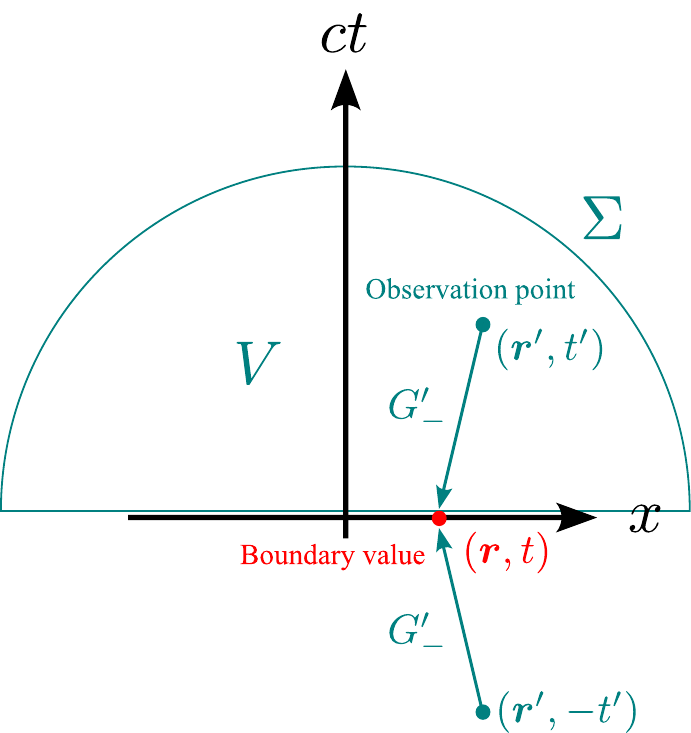}
    \caption{\textbf{Rayleigh's trick in spacetime}: the field at $(\boldsymbol{r}',t')$ is found through integrating an advanced Green function and its derivative (source point equals $(\boldsymbol{r}',t')$) over the boundary.  The Green function can be made to vanish on the $t=0$ boundary through including a mirrored (retarded) source outside $V$.  This ensures that the field at the observation point depends only on the field values on the $t=0$ boundary.
    }
    \label{fig:RayleighTrick}
\end{figure}

We begin with a simple example of our approach, considering the $(3+1)$ dimensional Klein--Gordon equation~\cite{KleinEqn,GordonEqn}: 
\begin{equation}\label{eq:KGEq}
	(\eta^{\mu\nu}\partial_\nu\partial_\mu+\rmM^2)\Psi=0
\end{equation}
with $\eta^{\mu\nu}={\rm diag}\{c^{-2},-1,-1,-1\}$ and $\rmM=\rmm\rmc/\hbar$. In the spirit of the diffraction theory, we start from the following identity:
\begin{equation}\label{eq:delta_identity}
\Psi({x'}^{\mu})=\int_V \delta^{(4)}(x^{\mu}-{x'}^{\mu})\Psi(x^{\mu})\,\rmd x\,\rmd y\,\rmd z\,\rmd t,
\end{equation}
where $\delta^{(4)}(x^{\mu}-{x'}^{\mu})$ is the four--dimensional delta function,  and $V$ is a four--dimensional volume containing the observation point $(x',y',z',t')$. Without loss of generality, let $(x',y',z',t')$ be a spacetime point in the future of $t$, i.e., $t'>t$. The four--dimensional volume $V$ is similarly chosen to be where $t'\geq t$. To make use of the identity (\ref{eq:delta_identity}), the delta function needs to be expressed in terms of the Green function for the Klein--Gordon equation (\ref{eq:KGEq}), which satisfies,
\begin{equation}
    (\eta^{\mu\nu}\partial_\nu\partial_\mu+\rmM^2)G({x}^{\xi}-{x'}^{\xi})=\delta^{(4)}({x}^{\xi}-{x'}^{\xi}).\label{eq:green-def}
\end{equation}
In addition, we need to transform the spacetime volume integral into a boundary integral.  To achieve both of these aims, we should express the integrand of (\ref{eq:delta_identity}) as a four--dimensional divergence.  By direct analogy with Kirchoff's approach to diffraction integrals, we observe that the integrand in \cref{eq:delta_identity} can---through application of Eq. (\ref{eq:green-def})---be expressed in as a four--divergence as follows,
\begin{equation}\label{eq:KGEq.div.GPsi}
-\eta^{\mu\nu}\partial_\mu\left[G\partial_\nu\Psi-\Psi\partial_\nu G\right]=\delta^{(4)}(x^{\mu}-{x'}^{\mu})\Psi.
\end{equation}
At this point we note that there are four different choices for the Green function.  These are the retarded and advanced Green functions (vanishing for $t-t'<0$ and $t-t'>0$ respectively), which are given by
\begin{equation}\label{eq:KGEq.G_pm}
  G_\pm(\bm R,\tau)=\frac{\rmM\rmc}{4\pi s}\Theta\left(\pm\tau-\frac R\rmc\right)J_1\left(\rmM s\right)-\frac{\rmc}{2\pi}\delta\left(\tau\mp\frac R\rmc\right),
\end{equation}
and the non--causal Green functions (these are non--zero both for $t>t' $ and $t<t'$), which are defined by,
\begin{multline}\label{eq:KGEq.G_pm.noncausal}
    G'_\pm(\bm R,\tau)=\frac{\mp\rmi\rmc}{(2\pi)^2s}\frac{\rmd}{\rmd s}\bigg[\Theta(s^2)\frac{\pm\rmi\pi}2 H_0^{(\frac{3\mp1}2)}(\mu s)\\+\Theta(-s^2)K_0\left(\mu \sqrt{-s^2}\right)\bigg],
\end{multline}
where in both \cref{eq:KGEq.G_pm,eq:KGEq.G_pm.noncausal} we have now separated the time and space coordinates and used $s^2=\rmc^2\tau^2-R^2$ with $\tau=t-t'$, $R=|\bm R|=|\bm r-\bm r'|$, $\bm r=(x,y,z)$ and $\bm r'=(x',y',z')$.  In these equations $\Theta$ is the Heaviside step function, $J_n$ the $n$--order Bessel function of the first kind, $H_n^{((3\mp1)/2)}$ is the $n$--order Hankel Function of the first or second kind (corresponding to $(3\mp1)/2=1,\ 2$), and $K_n$ is the $n$--order modified Bessel function of the second kind. Substituting \cref{eq:KGEq.div.GPsi} into \cref{eq:delta_identity} gives
\begin{equation}\label{eq:KGEq.div}
\Psi(\bm r',t')=-\int_V \eta^{\mu\nu}\partial_\mu\left[G\partial_\nu\Psi-\Psi\partial_\nu G\right]\rmd x\,\rmd y\,\rmd z\,\rmd t.
\end{equation}
Assuming the spacetime volume indicated in Fig. \ref{fig:RayleighTrick}, and applying the four--dimensional Gauss theorem to \cref{eq:KGEq.div}, the value of the wave at $\boldsymbol{r}',t'$ becomes equal to a boundary integral,
\begin{multline}\label{eq:KGEq.diffInt.proto}
\Psi(\bm r',t')=\int_{t=0} \frac1{\rmc^2}\left[G\partial_t\Psi-\Psi\partial_t G\right]\rmd x\rmd y\rmd z
\\
+\int_{\Sigma}  [\Psi\partial_\mu G-G\partial_\mu \Psi]\rmd^3 S^\mu,
\end{multline}
where the $\Sigma$ denotes the upper boundary shown in Fig.~\ref{fig:RayleighTrick}  and $\rmd^3 S^\mu$ denotes the infinitesimal surface element in four--dimensional space, along the upper boundary $\Sigma$.

To eliminate the temporal derivative term of $\Psi$ in \cref{eq:KGEq.diffInt.proto}, we can define the Green function as arising from two sources, one of which is outside of the spacetime integration region $V$, as shown in \cref{fig:RayleighTrick}.  Being outside of the integration region, this second source does not modify the left hand side of \cref{eq:KGEq.diffInt.proto}.  Mirroring these two sources in time, we can thus choose the Green function in \cref{eq:KGEq.diffInt.proto} to vanish on the $t=0$ boundary, defining
\begin{equation}\label{eq:KGEq_RGfunc}
    G_R(\bm r,\bm r';t,t')=G'_-(\bm R,t-t')-G'_-(\bm R,t+t') ,
\end{equation}
where $G'_-$ is the Green function defined in \cref{eq:KGEq.G_pm.noncausal}, which we refer to as the ``Rayleigh--Green function''.  As noted in the caption to \cref{fig:RayleighTrick}, this approach is formally identical to that used by Rayleigh in the theory of diffraction, where the $t=0$ boundary is, in that case a spatial boundary.  

As it stands, \cref{eq:KGEq.diffInt.proto} requires us to specify the value of the field $\Psi$ on \emph{both} the $t=0$ and the infinite future, $\Sigma$ spacetime surfaces.  In diffraction theory, the contribution of the analogous spatial boundary at infinity is eliminated using the Sommerfeld condition~\cite{born2013principles}, which is equivalent to the statement that all energy travels outwards to infinity.  We now give the boundary condition necessary to eliminate the contribution of the field on the boundary $\Sigma$ in spacetime to eliminate the upper boundary integral in \cref{eq:KGEq.diffInt.proto}.

Performing the Fourier transformation with respect to $3$D spatial coordinates and replacing the Green function in \cref{eq:KGEq.diffInt.proto} with the Rayleigh--Green function \barecref{eq:KGEq_RGfunc} yields
\begin{multline}\label{eq:KGEq.div.k.boundary}
\Psi(\bm k,t')=\frac1{\rmc^2}\bigg[\Psi(\bm k,t)\partial_t G_R(t-t';\bm k)
\\
-G_R(t-t';\bm k)\partial_t\Psi(\bm k,t)\bigg]_{t=0}^\infty,
\end{multline}
where $\Sigma$ is now the upper $t=\infty$ boundary on the square brackets.  Since, by definition the ``Rayleigh--Green function'' $G_{R}$ vanishes at $t=0$, \cref{eq:KGEq.div.k.boundary} simplifies to
\begin{multline}\label{eq:KGEq.div.k.boundary.eval}
    \Psi(\bm k,t')=-\frac1{\rmc^2}\Psi(\bm k,t)\partial_t G_R(t-t';\bm k)\bigg|_{t=0}
    \\
    +\frac1{\rmc^2}\bigg[\Psi(\bm k,t)\partial_t G_R(t-t';\bm k)
    \\
    -G_R(t-t';\bm k)\partial_t\Psi(\bm k,t)\bigg]_{t\to\infty}.
\end{multline}
The corresponding Fourier transformation of the Rayleigh--Green function \barecref{eq:KGEq_RGfunc}, $G_R(t-t';\bm k)$, in \cref{eq:KGEq.div.k.boundary.eval} is then given by
\begin{equation}\label{eq:KGEq_RGfunc.noncausal.k}
    G_R(t-t';\bm k)=-\frac{\rmi\pi}{\omega_{\bm k}}\left(\rme^{-\rmi\omega_{\bm k}|t-t'|}-\rme^{-\rmi\omega_{\bm k}|t+t'|}\right).
\end{equation}
Note we have here chosen the negative exponential, the positive sign also being admissible providing we modify the boundary condition in \cref{eq:SommerfeldConditionInSpacetime}, below.  Substituting \cref{eq:KGEq_RGfunc.noncausal.k} into \cref{eq:KGEq.div.k.boundary.eval}, we find that, in order to eliminate the second term corresponding to the upper boundary $t\to\infty$, the wave function must satisfy the following condition:
\begin{equation}\label{eq:SommerfeldConditionInSpacetime}
    \lim_{T\to\infty}\rmi\omega_{\bm k}\Psi(\bm k, T)+\partial_t\Psi(\bm k,T)=0,
\end{equation}
where $\omega_{\boldsymbol{k}}=c\sqrt{|\boldsymbol{k}|^2+M^2}>0$.  We observe that \cref{eq:SommerfeldConditionInSpacetime} is the spacetime analogue of the usual Sommerfeld condition discussed above, and is equivalent to the requirement that each Fourier component of the field evolves in time with a positive frequency, i.e. as $\exp(-{\rm i}\omega_{\boldsymbol{k}}t)$.  Therefore, applying condition \cref{eq:SommerfeldConditionInSpacetime}, the Rayleigh--Sommerfeld diffraction integral for the Klein--Gordon equation is given by the simple expression
\begin{equation}\label{eq:KGEq.diffInt.RS}
\Psi(\bm r',t')=-\frac1{\rmc^2}\int_{t=0} \Psi(\boldsymbol{r},0)\,\partial_t G_R(\boldsymbol{r},\boldsymbol{r}';t,t')\, \rmd x\,\rmd y\,\rmd z.   
\end{equation}
From our simple argument we can thus see that the propagator (for positive frequency fields) for the Klein--Gordon equation is given by
\begin{equation}
    K(\boldsymbol{r},\boldsymbol{r}';t,t')=-\frac1{\rmc^2}\partial_t G_R(\boldsymbol{r},\boldsymbol{r}';t,t').
\end{equation}
where the Green function is the ``Rayleigh--Green function'' defined in \cref{eq:KGEq_RGfunc}.

\section{The diffraction integral for the Dirac equation}

Having found the propagator from the Klein--Gordon equation using an analogous argument to Rayleigh's, we apply the idea to the Dirac equation~\cite{DiracEqn}:
\begin{equation}
(\rmi\gamma^\mu\partial_\mu-\rmM \mathbbm{1})\Psi=0,
\end{equation}
where $\mathbbm{1}$ is the identity matrix, the $\gamma^{\mu}$ are the four Dirac gamma matrices obeying $\{\gamma^{\mu},\gamma^{\nu}\}=2\eta^{\mu\nu}$, and $M$ is defined as in the previous section.  We again start from the delta function identity \cref{eq:delta_identity} and introduce the (matrix) Green function, obeying $(\rmi\gamma^\mu\partial_\mu-\rmM)G=\delta^{(4)}(x^{\nu}-{x'}^{\nu})$.  Similar to \cref{eq:KGEq.div.GPsi}, we find that
\begin{equation}\label{eq:DiracEq.div.GPsi}
\partial_\mu(-\rmi \bar{G}\gamma^\mu\Psi)=\delta^{(4)}(x^{\nu}-{x'}^{\nu})\Psi,
\end{equation}
where $\bar G$ denotes the Dirac adjoint of the Green function, i.e., $\bar{G}=\gamma^0 G^\dagger\gamma^0$ with $\dagger$ denoting the Hermitian conjugate. Again, as in the previous example, there are four different choices for the Green function.  By contrast here we choose the \emph{causal} (advanced) Green function:
\begin{equation}\label{eq:Dirac_G_proto}
    G=-(\rmi\gamma^\mu\partial_\mu+\rmM\mathbbm{1})G_-,
\end{equation}
where $G_-$ is the advanced Green function defined in \cref{eq:KGEq.G_pm}.  Integrating \cref{eq:DiracEq.div.GPsi} over the same spacetime domain as shown in \cref{fig:RayleighTrick}, i.e. $t>0$, we again obtain a Rayleigh--like diffraction integral for the Dirac equation:
\begin{equation}\label{eq:3+1DiracEq.diffraction_integral}
    \Psi(\bm r',t')=\int_{t=0} \frac\rmi\rmc \gamma^0 G^\dagger(\bm r-\bm r',t-t')\Psi(\bm r,t)\,\rmd x\,\rmd y\,\rmd z.
\end{equation}
In this case the integration over the upper boundary $\Sigma$ vanishes since the Green function itself vanishes there.  The propagator for the Dirac equation is therefore given by
\begin{equation}\label{eq:Dirac_propagator}
    K(\bm r-\bm r',t-t')=\frac\rmi\rmc\gamma^0 G^\dagger(\bm r-\bm r',t-t').
\end{equation}
In the special case of the $(1+1)$ dimensional Dirac equation we can choose $\gamma^0=\sigma_x$ and $\gamma^1=-\rmi\sigma_y$, the integral \barecref{eq:3+1DiracEq.diffraction_integral} reduces to one over a single spatial coordinate, $x$, and the advanced Green function appearing in \cref{eq:Dirac_G_proto} is given by (see appendix for derivation), 
\begin{equation}\label{eq:KGEq_RAGreen}
   G_{-}(x-x',\tau)=\frac\rmc2\Theta\left(-\tau-\frac{|x-x'|}{\rmc}\right)J_0\left(\rmM s\right),
\end{equation}
where $\tau$ was defined in the previous section.  Combining \cref{eq:KGEq_RAGreen,eq:Dirac_G_proto,eq:Dirac_propagator}, the propagator \barecref{eq:Dirac_propagator} for the 1+1 dimensional Dirac equation becomes
\begin{equation}\label{eq:Dirac_K}
    \begin{array}{cl}
         K=& -\frac12\left\{ \rmi\sigma_x \rmM\Theta\left(-\tau-\frac{|x-x'|}\rmc\right)J_0(\rmM s) \right.
         \\[5pt]
         &-\mathbbm{1}\Theta\left(-\tau-\frac{|x-x'|}\rmc\right)\frac{\rmM\rmc\tau}sJ_1(\rmM s)
         \\[5pt]
         &-\sigma_z\Theta\left(-\tau-\frac{|x-x'|}\rmc\right)\frac{\rmM(x-x')}sJ_1(\rmM s)
         \\[5pt]
         & -\frac1\rmc\mathbbm{1}\delta\left(-\tau-\frac{|x-x'|}\rmc\right)
         \\[5pt]
         &\left. +\sigma_z\frac1\rmc\delta\left(-\tau-\frac{|x-x'|}\rmc\right){\rm sign}(x-x') \right\}.
    \end{array}
\end{equation}
where $s=\sqrt{\rmc^2\tau^2-|x-x'|^2}$.  Note, in the special case when the mass is zero, the propagator \barecref{eq:Dirac_K} is simply a sum of two $\delta$ functions, indicating free propagation at the speed of light.  As we shall see, these $\delta$ functions arise from propagation along the light cone, which appear to have been overlooked in previous treatments of the Feynman checkerboard model.

\section{Comments on the Feynman checkerboard model}

\begin{figure}
    \centering
    \includegraphics[width=0.8\linewidth]{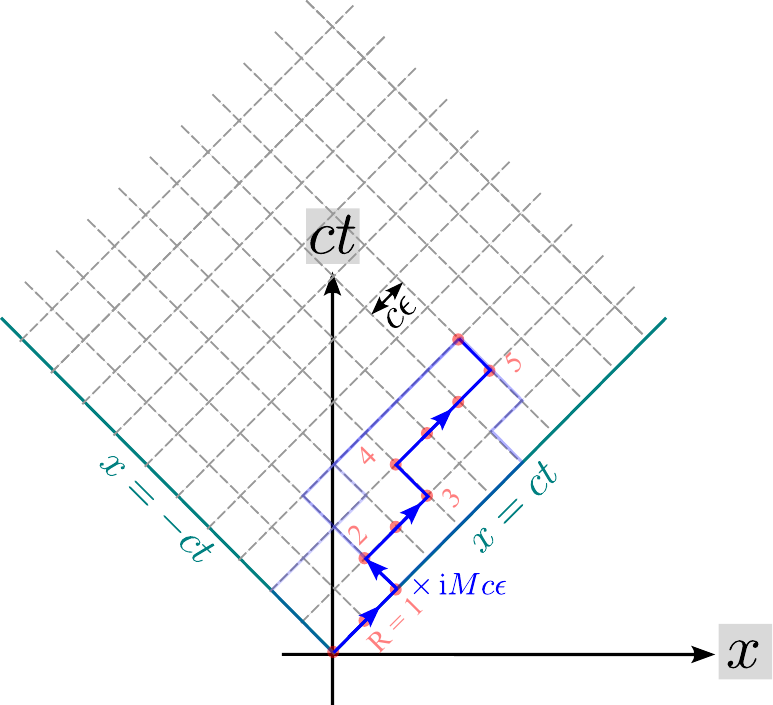}
    \caption{\textbf{The Feynman checkerboard model}: The propagator for the $(1+1)$ dimensional Dirac equation can be constructed according to Eq. (\ref{eq:FeynmanDef_K}), as a sum over paths, each travelling at the speed of light $x=\pm ct$.  Here we show one such path as the dark blue zigzag line (other contributions represented as faded blue lines), which has $R=5$ corners in this example.  This trajectory illustrates that the particle occasionally moves forward or backward at light speed, with the rate of interchange determined by the value of $M$, defined below \cref{eq:KGEq}.  At each corner the contribution to the propagator from the chosen path gains a factor of $\rmi M c\epsilon$.}
    \label{fig:FCM}
\end{figure}

The propagator for the Dirac equation was first discussed by Feynman \cite{FeynmanQM}, via the Feynman checkerboard model, which we have described in \cref{fig:FCM}.  This calculation of the propagator is notably different from the approach applied to the Schr\"{o}dinger equation.
In Feynman's original calculation of the propagator (\ref{eq:Dirac_K}), the particle can only move forward or backward in one dimension at the speed of light, $\rmc$. The particle occasionally reverses its direction at a rate proportional to the rest mass, $\rmm$. This generates a random walk, as sketched in \cref{fig:FCM}.  The corresponding wave function of the particle is assumed to change its amplitude by a factor $\rmi \rmM\rmc\epsilon$ after each reversal, where $\epsilon$ is the small time step into which the path is discretized (also shown in \cref{fig:FCM}).  Combining these assumptions, the propagator $K$ for the Dirac equation is equal to~\cite{FeynmanQM}
\begin{equation}\label{eq:FeynmanDef_K}
K=\underset{R}{\sum}N(R)(\rmi \rmM\rmc\epsilon)^R,
\end{equation}
where $R$ is the number of corners on the path illustrated in \cref{fig:FCM} (i.e. the number of times the trajectory switches between left--going and right--going), and $N(R)$ is the total number of paths with $R$ corners. The sum in the propagator \barecref{eq:FeynmanDef_K} can, after an evaluation of $N(R)$ using combinatorics, and an application of the Stirling formula for a large number of time steps~\cite{kull1999path}, be evaluated as
\begin{equation}\label{eq:FeynmanEva_K}
K=\sigma_xJ_0(\rmM s)+\rmi\frac{\rmc \tau}s J_1(\rmM s)+\rmi\frac{(x-x')}s\sigma_z J_1(\rmM s).
\end{equation}
The Pauli matrices in the propagator \barecref{eq:FeynmanEva_K} assume the representation of the gamma matrices in the $(1+1)$D Dirac equation with $\gamma^0=\sigma_x$ and $\gamma^1=-\rmi\sigma_y$.
We should also note that, in Feynman's model, all paths are confined inside the light cone as shown in \cref{fig:FCM}. To ensure zero propagation outside of the light cone, and confining ourselves to propagation into the future, $t'>t$, we should thus multiply \cref{eq:FeynmanEva_K} by the Heaviside step functions, $\Theta(-\tau)\Theta(s^2)=\Theta(-\tau-|x-x'|/c)$, as we have in \cref{eq:Dirac_K}.  

The propagator given in \cref{eq:FeynmanEva_K} is very closely related to the expression we found above, but neglects two terms, which are the delta functions in \cref{eq:Dirac_K}.  The three non--singular terms in propagator \barecref{eq:FeynmanEva_K}, derived from the Feynman checkerboard model in e.g.~\cite{kull1999path}, are identical to those in our result \barecref{eq:Dirac_K}, except for a normalization factor $-\rmi \rmM/2$ (we note that this factor is required to give the propagator the correct units of inverse length, and thus not alter the units of the field!).  We note also that our propagator \barecref{eq:Dirac_K} becomes a $\delta$ function when the evolution time, $\tau$ vanishes (as required to give back the original field distribution when taking the limit of zero time evolution).  In contrast, the propagator \barecref{eq:FeynmanEva_K} does not.  Furthermore, when the mass vanishes--- a case not obviously considered by Feynman, namely, $\rmm=0$,   the propagator \barecref{eq:FeynmanDef_K} is 1 on the light cone and 0 elsewhere.  But in this case, the Dirac equation is identical to the $(1+1)$D Maxwell equations in free space.  These $\delta$ terms in the propagator \barecref{eq:Dirac_K} simply mean that the wave propagates freely with a retarded time $\tau=|x-x'|/\rmc$.  This fact is excluded from the propagator\barecref{eq:FeynmanEva_K} which tends to unity when $\rmm=0$. In fact, the additional $\delta\left(\tau-|x-x'|/\rmc\right)$ terms originate from those paths on the light cone which have no corner in the Feynman checker board model, as illustrated as the teal line in \cref{fig:FCM}. These discrepancies arise because the Feynman checkerboard model is a discrete model.  For paths lying on the light cone, the amplitude computed from \cref{eq:FeynmanDef_K} is always 1, which, in fact we can understand as a Kronecker delta, equivalent to zero in the continuous limit, thus eliminating the required delta function terms from \cref{eq:FeynmanEva_K}.

In addition, in the $(3+1)$D Feynman checkerboard model, the $(3+1)$D  Dirac propagator \barecref{eq:Dirac_propagator} also differs from the one given in earlier works (e.g. Ref.~\cite{ord2023feynman}), as evident from the presence of Bessel and delta functions in \cref{eq:KGEq.G_pm}. Furthermore, when $\tau=0$, the propagator should be a delta function whereas the one in Ref.~\cite{ord2023feynman} is 1, which arises again from a Kronecker delta---the same issue discussed above, arising from the discrete nature of the Feynman checkerboard model.  We should emphasize that our propagator \cref{eq:Dirac_propagator} correct, being derived directly from the wave equation itself without relying on any particular model.

\section{Summary and conclusions}

We have developed an approach analogous to the Rayleigh--Sommerfeld diffraction integral, applied in spacetime, to find the propagators of both the Klein--Gordon and Dirac equations.  In particular we have compared our $(1+1)$D Dirac propagator~\barecref{eq:Dirac_K} with the propagator given by the Feynman checkerboard model~\cite{kull1999path,FeynmanQM}, finding that these previous results are missing terms, required to give the correct limit for both zero mass and the limit of zero time evolution.  We find that this problem results from the amplitude having been assigned as unity for those paths lying on the light cone, which, from our results should actually be singular, i.e. a contribution of $\delta(\tau-|x-x'|/\rmc)$ to the propagator in the continuous limit.  We found that the same problem is present in the $(3+1)$ dimensional propagator for the Dirac equation and this that previous propagators derived from the $(3+1)$D Feynman checkerboard model Ref.~\cite{ord2023feynman} have the same problem in the zero mass and null evolution limits.  Not only does our method provide a means to check and correct these previous results, but gives a new way to derive the time evolution of wave fields.

\section*{Acknowledgements}
Mingjie thanks the China Scholarship Council for financial support. SARH thanks the Royal Society and TATA for financial support (RPG-2016-186), as well as support through EPSRC program grant ``Next generation metamaterials: exploiting four dimensions'' EP/Y015673/1.

\section{Appendix}

Here we derive the Green function for $(1+1)$D Klein--Gordon equation.
The Green function should satisfy
\begin{equation}
    \left(\frac1{\rmc^2}\partial^2_t-\partial^2_x+\rmM^2\right)G(x-x',t-t')=\delta(x-x')\delta(t-t')
\end{equation}
which can, using the Fourier transformation, be solved as
\begin{equation}\label{eqAPP:G_1DKleinEq}
    G=\frac{-1}{4\pi^2}\iint^{+\infty}_{-\infty}\frac{\exp(\rmi(k(x-x')-\omega(t-t')))}{\frac{\omega^2}{\rmc^2}-k^2-\rmM^2}\rmd \omega\rmd k .
\end{equation}
There are two singular points $\omega=\pm\rmc\sqrt{k^2+\rmM^2}$ on the integral path $k={\rm constant}$. To avoid them, we should choose a contour going clockwise over (anti--clockwise under) both poles to give the retarded (advanced) Green function. Integrate \cref{eqAPP:G_1DKleinEq} over $\omega$ and replace $k$ with $\rmM \sinh{\eta}$, giving
\begin{equation}\label{eqAPP:Gpm}
    G_\pm=\frac{\Theta(\pm (t-t'))\Theta(s^2)}{2\pi}\int^\infty_{-\infty}\sin(\rmM s \cosh(\eta))\rmd\eta .
\end{equation}
Using the integral representation of Bessel function~\cite{BesselTreatise}, \cref{eqAPP:Gpm} yields \cref{eq:KGEq_RAGreen}.

\bibliography{Ref.bib}
\end{document}